\documentstyle[12pt]{article}

\begin{document}
\setcounter{page}{1} \pagestyle{plain} \vspace{1cm}
\begin{center}
\Large{\bf Minimal Length and Generalized Dirac Equation }\\
\small \vspace{1cm}
{\bf Kourosh Nozari}\quad and \quad {\bf Mojdeh Karami }\\
\vspace{0.5cm} {\it Department of Physics,
Faculty of Basic Science,\\
University of Mazandaran,\\
P. O. Box 47416-1467,
Babolsar, IRAN\\
e-mail: knozari@umz.ac.ir}
\end{center}
\vspace{1.5cm}
\begin{abstract}
Existence of a minimal observable length which has been indicated by
string theory and quantum gravity, leads to a modification of Dirac
equation. In this letter we find this modified Dirac equation and
solve its eigenvalue problem for a free particle. We will show that
due to background spacetime fluctuation, it is impossible to have
free particle in Planck scale.
\end{abstract}
\newpage
\section{Introduction}
The problem of reconciling Quantum Mechanics with General Relativity
is one of the task of modern theoretical physics which, until now,
has not yet found a consistent and satisfactory solution. One of the
most interesting consequences of this unification is that in quantum
gravity there exists a minimal observable distance on the order of
the Planck distance, $ l_{P} =\sqrt{\frac{G\hbar}{c^3}}\sim
10^{-33}cm$, where $G$ is the Newton constant. The existence of such
a fundamental length is a dynamical phenomenon due to the fact that,
at Planck scale, there are fluctuations of the background metric,
i.e. a limit of the order of Planck length appears when quantum
fluctuations of the gravitational field are taken into account. In
the language of string theory one can say that a string cannot probe
distances smaller than its length. The existence of such minimal
observable length which is motivated from string theory[1-5], loop
quantum gravity[6], Non-commutative geometry[7] and black hole
physics[8,9], leads to a generalization of Heisenberg uncertainty
principle to incorporate gravitational induced uncertainty from very
beginning. This generalized uncertainty principle can be written as
\begin{equation}
\label{math:1.1} \Delta x\geq \frac{\hbar}{\Delta p} +
\alpha'l_{P}^2\frac{\Delta p}{\hbar}.
\end{equation}
At energy much below the Planck mass, $ m_{P} =\sqrt{\frac{\hbar
c}{G}}\sim 10^{19}GeV/c^2$, the extra term in equation (1) is
irrelevant and the Heisenberg uncertainty relation is recovered,
while, as we approach the Planck energy, this term becomes relevant
and, as has been said, it is related to the existence of minimal
observable length. Now the generalized commutation relation becomes,
\begin{equation}
\label{math:1.2} [x,p]=i\hbar(1+\beta' p^2),
\end{equation}
where $\beta^{\prime}=\alpha^{\prime}\frac{l^{2}_{p}}{\hbar^2}$.
This feature constitutes a part of the motivation to study the
effects of this modified algebra on various observable. Consequences
of such a gravitational uncertainty principle (GUP), has been
studied extensively[10-25]. Here we proceed some more step in this
direction. We consider the effect of GUP on definition of momentum
operator and then we obtain a generalized Dirac equation and solve
the corresponding eigenvalue problem. We will show that it is
impossible to have free particle in Planck scale. Comparison between
our results and the results of ordinary relativistic quantum
mechanics shows a considerable departure when one approaches extreme
quantum gravity limit. We choose $\hbar=c=1$. In this situation
$\beta^{\prime}=l^{2}_{p}=\frac{1}{m^{2}_{P}}$. Latin index $i$,
takes the values $1,2,3$ and Greek index $\mu$ takes the values
$0,1,2,3$. Index $0$ stands for temporal part of 4-vectors.

\section{Generalized Dirac Equation}
Using GUP as equation (1), one can generalize usual momentum
operator $ p_{op} = -i \frac{\partial}{\partial x}$ to the following
form[19,20]
\begin{equation}
\label{math:2.1} p^{(GUP)}_{op}=-i\Bigg(1-\beta' \Big(
\frac{\partial}{\partial x}\Big)^{2}\Bigg)\frac{\partial}{\partial
x},
\end{equation}
or in three-dimensional form,
\begin{equation}
\label{math:2.2} \vec{p}=
-i\vec{\nabla}\Big(1-\beta^{\prime}\vec{\nabla}^{2}\Big).
\end{equation}
To satisfy Lorentz invariance, this equation can be written as
\begin{equation}
\label{math:2.3}
p_{i}=-i\Big[1-\beta'(\partial_{i})^2\Big]\partial_{i}.
\end{equation}
The square of this operator is,
\begin{equation}
\label{math:2.4}
p^{2}=p_{i}p^{i}=\bigg(-i\Big[1-\beta'(\partial_{i}\partial^{i})\Big
]\partial_{i}\bigg)\bigg(-i\Big[1-\beta'(\partial^{i}\partial_{i})\Big]\partial^{i}\bigg),
\end{equation}
which up to first order in $\beta^{\prime}$ simplifies as
\begin{equation}
\label{math:2.5}
p^{2}\simeq-\bigg[\partial_{i}\partial^{i}-2\beta'\Big(\partial^{i}\partial_{i}\Big)\Big(\partial^{i}\partial_{i}\Big)
\bigg].
\end{equation}
Now the Dirac equation[26],
\begin{equation}
\label{math:2.6} \Big(\gamma^{\mu}p_{\mu}-m\Big)\psi
=\Big(i\gamma^{\mu}\partial_{\mu} + m\Big)\psi =0,
\end{equation}
which can be written as
\begin{equation}
\label{math:2.7}i\partial_{0}\psi=
-\gamma_{0}(i\gamma^{i}\partial_{i}+m)\psi,
\end{equation}
should be written in generalized form. For this purpose, in addition
to generalization of momentum operator, we should consider the
generalization of energy operator also. Usually one considers the
energy operator as $E=i\partial_{0}$. As a consequence of
generalized uncertainty principle, one can define generalized
frequency as[27]
\begin{equation}
\label{math:2.8} \tilde{\omega}= E-\beta'E^3=E(1-\beta' E^{2}),
\end{equation}
where a numerical factor of $1/3$, which is irrelevant to our
purposes has been omitted. Using the energy mass shell condition $
p^2+m^2=E^2$, one finds
\begin{equation}
\label{math:2.9} \tilde{E} =
E\Big(1-\beta'(p^2+m^2)\Big)=E\Big(1-\beta' p^2- \beta'm^2\Big).
\end{equation}
Now the generalized Dirac equation can be written as
\begin{equation}
\label{math:2.10}i\partial_{0}\tilde{\psi}=-\gamma_{0}\Big(i\gamma^{i}
\partial_{i}+m \Big)(1- \beta'(p_{i}p^{i})- \beta' m^2)\tilde{\psi}.
\end{equation}
Since $p_{i}p^{i}$ is given by equation (7), we find the following
form for generalized Dirac equation
\begin{equation}
\label{math:2.11} i\partial_{0}\tilde{\psi}=
-\gamma_{0}\Bigg(i\gamma^{i}\partial_{i} +m
\Bigg)\Bigg[1+\beta'\Bigg(\partial_{i}\partial^{i} -m^2\Bigg)
\Bigg]\tilde{\psi},
\end{equation}
up to first order in $\beta^{\prime}$. In this equation
$\tilde{\psi}$ is generalized Dirac spinor which will be calculated.
Note that this equation has Lorentz invariance evidently. To solve
this eigenvalue equation, first we try to write it in a form which
be comparable easily with usual Dirac equation, (9). For this
purpose, the right hand side of equation (13) can be written as,
$$-\gamma_{0}\Big(i\gamma^{i}\partial_{i}+m\Big)
\Big(1+\beta'\partial^{i}\partial_{i}-\beta' m^2\Big)\tilde{\psi},$$
\begin{equation}
\label{math:2.12}
=-\gamma_{0}\Big(i\gamma^{i}\partial_{i}+i\beta'\gamma^{i}(\partial^{i}\partial_{i})\partial_{i}-\beta'm^2i\gamma^{i}\partial_{i}
+m+\beta'm\partial_{i}\partial^{i}-\beta'm^3\Big)\tilde{\psi}.
\end{equation}
Therefore, equation (13) can be re-written as
\begin{equation}
\label{math:2.10}i\partial_{0}\tilde{\psi}=-\gamma_{0}\Big(i\gamma^{i}\partial_{i}+i\beta'\gamma^{i}(\partial^{i}\partial_{i})\partial_{i}-\beta'm^2i\gamma^{i}\partial_{i}
+m+\beta'm\partial_{i}\partial^{i}-\beta'm^3\Big)\tilde{\psi},
\end{equation}
which leads to
\begin{equation}
\label{math:2.13} \Big[E +[\beta\beta'
m(p^2-m^2)]\Big]\tilde{\psi}=\Big[\Big({\bf{\alpha}}.{\bf
p}\Big)\Big(1-\beta'(\nabla^2+m^2)+\beta m\Big)\Big]\tilde{\psi}.
\end{equation}
In comparison with equation (9), here we define a "generalized
potential"
\begin{equation}
\label{math:2.14} V_{0}=\beta'm(p^2-m^2)
\end{equation}
and "generalized momentum"
\begin{equation}
\label{math:2.15} \tilde{\Pi}=p\Big(1-\beta'(\nabla^2+m^2)\Big).
\end{equation}
Note that when $\beta^{\prime}\rightarrow 0$, one finds
$V_{0}\rightarrow 0$ and this shows that $V_{0}$ can be interpreted
as a potential related to quantum gravitational effects. In the
presence of this generalized potential the particle is no longer
free. Therefore in Planck scale there is no free particle due to
quantum fluctuations of background spacetime. Now the generalized
Hamiltonian becomes,
\begin{equation}
\label{math:2.16} \tilde{H}=({\bf
\alpha.p})\Big(1-\beta'(\nabla^2+m^{2})\Big)+\beta m
\end{equation}
and therefore generalized Dirac equation can be written as
\begin{equation}
\label{math:2.17}(E+\beta V_{0})\tilde{\psi}=\tilde{H}\tilde{\psi}.
\end{equation}
To solve this generalized Dirac equation for "free particle", we consider the following solution
\begin{equation}
\label{math:2.18}\tilde{\psi}(z,t)=\tilde{\psi}(z)e^{-i\varepsilon
t},
\end{equation}
where $\varepsilon=\lambda(E+V_{0})$ and we have assumed that the
particle is moving in $z$-direction. $\lambda=+1$ or $-1$ is related
to particle and anti-particle spectrum respectively. We consider the
case $\lambda=+1$ in which follows. Now, regarding to the
4-dimensional nature of spinors, we have
\begin{equation}
\label{math:2.19}\tilde{\psi}(z)=\pmatrix{\varphi\cr\chi}=\pmatrix{\psi_{1}\cr\psi_{2}\cr\psi_{3}\cr\psi_{4}}.
\end{equation}
Using operator form of $\tilde {H}$ as equation (19), we find
\begin{equation}
\label{math:2.19}\Bigg[E\pmatrix{1\quad
0\cr0\quad1}+V_{0}\pmatrix{1\quad 0\cr0\quad-1}\Bigg]
\pmatrix{\varphi\cr\chi}=\pmatrix{{\bf 0}\quad{\bf
\sigma}\cr{\bf\sigma} \quad {\bf 0}}.{\bf
p}\Bigg(1-\beta'(\nabla^2+m^2)\Bigg)\pmatrix{\varphi\cr\chi}
+m\pmatrix{1 \quad 0\cr0 \quad-1}\pmatrix{\varphi\cr\chi}.
\end{equation}
We should find eigenvalues and eigenvectors of this equation.
Suppose that
\begin{equation}
\label{math:2.20}\varphi(z)=\varphi_{0}e^{ip(1-\beta'p^2)z}
\end{equation}
and
\begin{equation}
\label{math:2.21}\chi(z)=\chi_{0}e^{ip(1-\beta' p^2)z}.
\end{equation}
Two points should be notified here: first note that we have
considered the generalized form of momentum operator in these
equations. Secondly, existence of minimal length destroys the notion
of locality. As Kempf has shown, in this situation one should
consider maximally localized states[25]. Therefore (24) and (25) are
maximally localized states. After substituting these two supposed
solutions in equation (23), some simple algebra gives
$$(E+V_{0})\varphi_{0}=(\sigma_{z}.p)\big[1-\beta'(2p^2+m^2)\big]\chi_{0}+m\varphi_{0}$$
\begin{equation}
\label{math:23}(E-V_{0})\chi_{0}=(\sigma_{z}.p)\big[1-\beta'(2p^2+m^2)\big]\varphi_{0}-m\chi_{0},
\end{equation}
where leads to
\begin{equation}
\label{math:2.24}\chi_{0}=\frac{(\sigma_{z}.p)\bigg(1-\beta'(2p^2+m^2)\bigg)}{E-V_{0}+m}\varphi_{0}.
\end{equation}
Now suppose that
\begin{equation}
\label{math:2.25}\varphi_{0}=\pmatrix{1\cr0},
\end{equation}
then it follows that
\begin{equation}
\label{math:2.25}\chi_{0}=\frac{(\sigma_{z}.p)\bigg(1-\beta'(2p^2+m^2)\bigg)}{E-V_{0}+m}\pmatrix{1\cr0}.
\end{equation}
To find energy spectrum of the system one should evaluate the
following $4\times4$ determinant
\begin{equation}
\label{math:2.26} \left|\begin{array}{cc}
  E+V_{0}-m & -(\sigma_{z}.p)\bigg(1-\beta'(2p^2+m^2)\bigg)\\
  -(\sigma_{z}.p)\bigg(1-\beta'(2p^2+m^2)\bigg) & E-V_{0}+m
  \end{array} \right|=0.
\end{equation}
Using the identity
\begin{equation} \label{2.27} ({\bf \sigma}.{\bf
A})({\bf \sigma}.{\bf B})={\bf A}.{\bf B}+i{\bf \sigma}.({\bf
A}\times{\bf B})
\end{equation}
a simple calculation gives
\begin{equation}
\label{math:2.27}
 E^2=p^2+m^2+\Big[2\beta'(2p^4-p^2m^2+m^4]\Big]
\end{equation}
up to first order in $\beta^{\prime}$. The term in curly bracket
gives the correction due to quantum gravitational effects. In usual
quantum mechanics, $\beta^{\prime}\ll 1$, and therefore one recovers
the well-known result for relativistic free particle. In extreme
quantum gravity limit when $\beta^{\prime}\sim 1$, there is a
considerable departure from usual picture.\\
Now the time-dependent wave function of the system becomes,
\begin{equation}
\label{math:2.28} \tilde{\psi}=Ne^{i[p(1-\beta'
p^2)z-\lambda(E+V_{0})t]}
\pmatrix{\varphi\cr\frac{(\sigma_{z}.p)\bigg(1-\beta'(2p^2+m^2)\bigg)}
{E+m-V_{0}}\varphi},
\end{equation}
or equivalently
\begin{equation}
\label{math:2.29} \tilde{\psi}=Ne^{i[p(1-\beta'
p^2)z-\lambda(E+V_{0})t]}
\pmatrix{\pmatrix{1\cr0}\cr\frac{(\sigma_{z}.p)(1-\beta'(2p^2+m^2))}{E+m-V_{0}}\pmatrix{1\cr0}}.
\end{equation}
The normalization constant can be calculated using orthogonality
condition
\begin{equation}
 \label{math:2.30}
\int\tilde{\psi}_{\lambda,p}^{\dag}(z,t)\tilde{\psi}_{\lambda,p}(z,t)d^{3}z=\delta_{\lambda,\lambda'}\delta(p-p'),
\end{equation}
which leads to
\begin{equation}
 \label{math:2.31}
N^2\Bigg(\varphi^{\dag}\varphi+\varphi^{\dag}\Big(\frac{(\sigma_{z}.p)(1-\beta'(2p^2+m^2))}
{E+m-V_{0}}\Big)^2\varphi\Bigg)=1
\end{equation}
or
\begin{equation}
 \label{math:2.32}
 N={\sqrt\frac{(E+m-V_{0})^2}{(E+m-V_{0})^2+\Bigg((\sigma_{z}.p)(1-\beta'(2p^2+m^2))\Bigg)^2}}.
\end{equation}
In terms of explicit form of $V_0$ and $\tilde{\Pi}$, and using
equation (31), this normalization factor becomes
\begin{equation}
\label{math:2.33}
 N={\sqrt\frac{p^2+2m^2+4\beta^{\prime}p^4+2Em}{2m^2
+2p^2-2\beta^{\prime}p^2m^2+2Em}}.
\end{equation}
Now the complete wave function for "free particle" in GUP is given
by
\begin{equation}
\label{math:2.34}
\tilde{\psi}_{p,\lambda}(z,t)={\sqrt\frac{p^2+2m^2+4\beta^{\prime}p^4+2Em}{2m^2
+2p^2-2\beta^{\prime}p^2m^2+2Em}}
\pmatrix{1\cr0\cr\frac{(\sigma_{z}.p)(1-\beta'(2p^2+m^2))}{E+m-V_{0}}\cr0}
e^{i[p(1-\beta'p^2)z-\lambda(E+V_{0})t]}
\end{equation}
for $\lambda=\pm1$. In the limit of $\beta^{\prime}\rightarrow 0$,
this equation reduces to usual Dirac spinor for free particle[26].
Now one should consider Helicity. In relativistic quantum mechanics,
Helicity is defined as[26]
\begin{equation}
\label{math:2.39} \Lambda_{s}={\bf{\Sigma}}.{\bf{\hat p}},
\end{equation}
where
\begin{equation}\label{math:2.40}
{\bf{\Sigma}}=
 \left(\begin{array}{cc}
   {\bf{\sigma}} &{\bf 0}\\
 {\bf 0} &{\bf{\sigma}}
  \end{array}\right),
\end{equation}
and ${\bf{\hat p}}=\frac{{\bf p}}{p}$. When one considers the
generalized form of momentum operator, the ${\bf{\hat p}}$ unit
vector does not change and therefore, Helicity operator suffers no
change in GUP. It is straightforward to show that generalized
Hamiltonian and generalized momentum operator commute with Helicity
operator, i.e. $[\tilde{H},\Lambda_{s}]=0$ and
$[\tilde{\Pi},\Lambda_{s}]=0$. Now we should consider this further
degree of freedom in complete wave function. Since we have
considered the $z$-direction as our preferred direction, there are
two possible eigenvalues for Helicity operator: $\pm 1/2$. Therefore
the complete Dirac spinor in the framework of GUP for "free
particle" is given by
\begin{equation}
\label{math:2.44}
\tilde{\psi}_{p,\lambda,+\frac{1}{2}}={\sqrt\frac{p^2+2m^2+4\beta^{\prime}p^4+2Em}{2m^2
+2p^2-2\beta^{\prime}p^2m^2+2Em}}
\pmatrix{\pmatrix{1\cr0}\cr\frac{(\sigma_{i}.p)(1-\beta'(2p^2+m^2))}{E+m-V_{0}}\pmatrix{1\cr0}}
e^{i[p(1-\beta'p^2)z-\lambda(E+V_{0})t]},
\end{equation}
for positive Helicity and
\begin{equation}
\label{math:2.46}
\tilde{\psi}_{p,\lambda,-\frac{1}{2}}={\sqrt\frac{p^2+2m^2+4\beta^{\prime}p^4+2Em}{2m^2
+2p^2-2\beta^{\prime}p^2m^2+2Em}}
\pmatrix{\pmatrix{0\cr1}\cr\frac{(\sigma_{i}.p)(1-\beta'(2p^2+m^2))}{E+m-V_{0}}\pmatrix{0\cr1}}
e^{i[p(1-\beta'p^2)z-\lambda(E+V_{0})t]},
\end{equation}
for negative Helicity. The orthogonality condition,
\begin{equation}
 \label{math:2.47}
\int\tilde{\psi}_{p_{z},\lambda,s_{z}}^{\dag}(z,t)\tilde{\psi}_{p'_{z},\lambda,s'_{z}}(z,t)d^{3}x=
\delta_{\lambda,\lambda'}\delta_{s_{z},s'_{z}}\delta(p_{z}-p'_{z})
\end{equation}
is satisfied properly.
\section{Summary}
Gravity induces uncertainty. Considering this additional
gravitational induced uncertainty, usual uncertainty principle of
Heisenberg should be modified. This fact leads to generalized
uncertainty principle. An immediate consequence of this generalized
uncertainty principle is generalization of momentum operator. This
generalized form of momentum operator should leads us to a
generalized Dirac equation. In this paper we have found this
generalized Dirac equation and have solved its eigenvalue problem.
We have shown that in Planck scale, due to quantum fluctuation of
background spacetime, there is no free particle. This have led us to
define a quantum gravitational generalized potential. Our
calculation shows that Helicity operator do not changes in GUP and
therefore complete generalized spinor of "free particle" has been
found systematically. Comparison between our results with usual
result of relativistic particle shows that in extreme quantum
gravity limit, $\beta^{\prime}\rightarrow 0$, the departure from
ordinary results are considerable.

\end{document}